# Electrically Thick Fabry-Pérot Omega Bianisotropic Metasurfaces as Virtual Anti-Reflective Coatings and Non-Local Field Transformers

Sherman W. Marcus and Ariel Epstein

Andrew and Erna Viterbi Faculty of Electrical Engineering, Technion - Israel Institute of Technology, Haifa 3200003, Israel

Omega bianisotropic metasurfaces (OBMS) provide wave-control capabilities not previously available with Huygens' metasurfaces (HMS). These enhanced capabilities derive from the additional degree of freedom provided by OBMS, and are based on analyses of zero-thickness surfaces with abstract surface impedance properties. However, the design of practical metasurfaces has proven tedious. Herein, we propose a novel, easily designed structure to realize OBMSs. Extending our previous work, we show that an asymmetrical cascade of two judiciously engineered Fabry-Perot (FP) etalons could form an OBMS meta-atom to provide desired wave control capabilities. Implementing this FP-OBMS for anomalous refraction, we show that bianisotropy effectively produces a virtual anti-reflective coating over a HMS, leading to the OBMS efficiency enhancement. This intriguing observation, backed by an approximate closed-form solution, provides an original physical interpretation of the mechanism underlying perfect anomalous refraction, and is used to explain for the first time differences in the angular response of OBMS in comparison to HMS. Implementing the FP-OBMS for the more intricate functionality of beam-splitting, we show that the FP-OBMS are capable of non-local excitation of surface waves required to this end, despite being electrically thick. These results, verified via full-wave computations, demonstrate the ability of the proposed physical structure to meticulously reproduce the scattering properties of ideal (abstract) zero-thickness OBMS, thereby paving the path to practical realization of wave transmission with exotic functionalities, some of which have never before been associated with a physical structure.

## I. INTRODUCTION

Metasurfaces have been studied in recent years for the purpose of producing unique reflected and/or transmitted fields. These metasurfaces consist of sub-wavelength elements known as "meta-atoms" which interact with the ambient field to produce the desired field effects. Huygens' metasurfaces (HMS) are characterized by their surface electric impedance $Z_{se}$ and surface magnetic admittance $Y_{sm}$ which determine the electric and magnetic current densities induced on the surface [1-11]. The two surface parameters $Z_{se}$ and $Y_{sm}$ have been shown sufficient to efficiently produce such phenomena as anomalous refraction for moderate angles of incidence and transmission [12]. However, for wide-angle beam deflection, a non-negligible reflected wave must be present to compensate for the wave impedance mismatch [7,13].

The failure of the HMS to transmit all the energy in the desired direction can be traced to the fact that it provides only two degrees of freedom in its surface characteristics: $Z_{se}$ and $Y_{sm}$ [7,13,14]. This has been overcome with the development of omega bianisotropic metasurfaces (OBMS) which possess an additional degree of freedom: a magneto-electric coefficient $K_{em}$ that couples the electric and magnetic surface current densities [8,15-17]. It has been shown that for an existing ambient electromagnetic field, an OBMS can in principle be designed to interact with this ambient field to produce any field of interest as long as the new field satisfies Maxwell's equations, and the power flow across the surface is locally continuous [16,18-22].

These conclusions, as well as the general wave-control capabilities of both the HMS and the OBMS, are based on analyses for ideal, zero-thickness surfaces characterized by surface susceptibilities that affect surface electric and magnetic currents. In order to synthesize an actual, physically realizable structure to emulate the effect of the abstract surface, physically realizable characteristics of the abstract system must be extracted, and these real characteristics mimicked in the actual structure. Physically realizable characteristics for this purpose are, for instance, the local values of the transmission coefficient $T(x)$ and the reflection coefficient $R(x)$ for normal incidence at each value of $x$ along the surface [10,23]. Whether this is accomplished can usually be determined only by full-wave scattering simulations [10]. The responses as a function of the element characteristics (size, shape, material), and their relationship to the metasurface properties $Z_{se}$, $Y_{sm}$ and $K_{em}$, would then be tabulated in a look-up table, would require multiple iterations, and would often require refinement through optimization [24-29]. Finally, entries in this table would be chosen which provided the closest value to the required $Z_{se}$, $Y_{sm}$ and $K_{em}$ for each meta-atom; this could often result in a tedious and less-than-precise design procedure.

In previous work, we have tackled this realization problem for HMSs using an original physical structure, based on a periodic array of parallel plate waveguides [30]. A sample structure that is being proposed here for the OBMS is shown in Fig. 1 and, like the previously-developed HMS structure, is filled with two cascaded dielectric layers. For the HMS case, these layers were arranged symmetrically about a centerline of the structure (i.e. $w_1=w_5$, $w_2=w_4$), consistent with previous suggestions that HMS structures should exhibit such symmetry [6,24]. The widths of the layers were adjusted in a Fabry-Pérot manner to provide the desired magnitude and phase of $R(x)$ and $T(x)$. These adjustments were made by varying the widths of the dielectric layers and the width of the space between them, two parameters corresponding to the two degrees of freedom, $Z_{se}$ and $Y_{sm}$, available for the HMS.

This Fabry-Pérot HMS (FP-HMS) was designed to anomalously refract a plane wave at oblique incidence $\theta_{inc}$ into a transmitted plane wave normal from the surface. However, as indicated previously, the HMS generally produces an unwanted specularly reflected wave in addition to the desired transmitted wave [7,10,13]. This is the case both for the zero-thickness implementation of the HMS and for the Fabry-Pérot implementation described in [30].

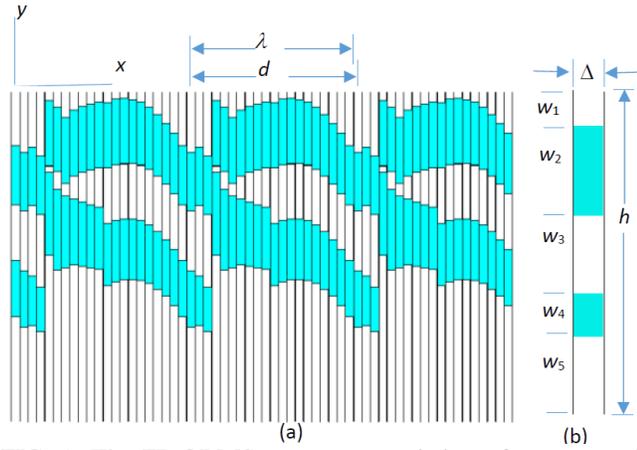

FIG. 1. The FP-OBMS structure consisting of an array of parallel plate waveguides containing a dielectric $\varepsilon_r$=16 (colored region) and air (white space). (a) Three periods of a structure to produce the $T(x)$ and $R(x)$ of Eq. (36) for anomalous refraction, $\theta_{inc}$=80°, $\theta_{trans}$=0, $h$=2$\lambda$, $N_{wg}$=18, $d$=1.0154$\lambda$ is the period (see Section IV). (b) A single meta-atom (parallel plate waveguide) displaying the widths $w_i$ of each layer.

In contrast, the zero-thickness OBMS has been shown capable of anomalous refraction, anomalous reflection, beam-splitting, and more advanced field transformations, without the presence of unwanted scattering [16-22,28, 29,31,32]. In view of these enhanced capabilities, and the previously discussed advantages of FP-HMSs, a method was sought for applying the Fabry-Perot structure to omega bianisotropic metasurfaces. In contrast to the FP-HMS, it would be envisioned that for such an "FP-OBMS", the dual layer geometry would be *asymmetric* (as shown in Fig. 1) [14,16,18,32-34], and would be established by adjusting *three* parameters representing the three degrees of freedom $Z_{se}$, $Y_{sm}$ and $K_{em}$.

In this paper, we develop a suitable semi-analytical methodology to attain these objectives, providing an efficient and reliable tool for generating physical FP-OBMS designs, and mimicking with high fidelity the performance of the abstract zero-thickness entities. The advantages of the FP-OBMS are demonstrated with the aid of two distinct applications: anomalous refraction and beam splitting. With anomalous refraction, we demonstrate that our proposed easily-designed structure can produce the perfect efficiency that is unattainable with HMS, and provide an additional analysis that reveals the underlying mechanism by which the OBMS removes the unwanted reflection. We utilize this analysis to explain for the first time (to the best of our knowledge) differences in the angular response of OBMS in comparison to HMS. For beam splitting we demonstrate that the required unconventional non-local excitation of auxiliary surface waves can be produced by our FP-OBMS despite being electrically thick, thereby allowing it to optimally perform this task. These results, confirmed via full-wave simulations, provide new physical insight into the performance of OBMSs, as well as indicate FP-OBMS configurations to be viable constructs for implementation of intricate field transformations.

## II. SYNTHESIS

In this section, the methodology will be discussed for designing a realistic thick structure that mimics an abstract zero-thickness OBMS. This is accomplished by generalizing the previously-developed FP-HMS structure and design procedure [30] to allow synthesis of the FP-OBMS.

The dielectric layers within each parallel plate waveguide (Fig. 1(a)) are required to produce specific local reflection and transmission coefficients $R(x)$ and $T(x)$ for normal incidence onto the waveguide located at $x$. These values of $R(x)$ and $T(x)$ are those which produce the desired wave-control effect for a zero-thickness structure. In the sub-sections below, the method of determining the $R(x)$ and $T(x)$ is detailed, and the method for realizing these $R(x)$ and $T(x)$ is described. For reasons that will be clarified in Section II-B, we restrict ourselves herein to transverse magnetic (TM) polarization.

### A. Local Transmission and Reflection Coefficients

The boundary conditions across the zero-thickness OBMS are expressed as "OBMS sheet transition conditions" (OB-STC) [8,15,16]. For TM polarization, these may be written in terms of the field components parallel to the surface:

$$\frac{1}{2}[E_x^+ + E_x^-] = Z_{se}(x)(H_z^+ - H_z^-) - K_{em}(x)(E_x^+ - E_x^-), \quad (1)$$

$$\frac{1}{2}[H_z^+ + H_z^-] = Y_{sm}(x)(E_x^+ - E_x^-) + K_{em}(x)(H_z^+ - H_z^-), \quad (2)$$

where $Z_{se}(x)$, $Y_{sm}(x)$ and $K_{em}(x)$ are the surface electric impedance, surface magnetic admittance and magneto-electric coefficient, respectively, at the location $x$; the $z$-components of the $H$-field along the top and the bottom of the surface are $H_z^+(x)$ and $H_z^-(x)$, respectively; and the $x$- components of the *E*-field along the top and the bottom of the surface are $E_x^+(x)$ and $E_x^-(x)$, respectively (see Fig. 2). For TM polarization, $H_x=H_y=E_z=0$ everywhere.

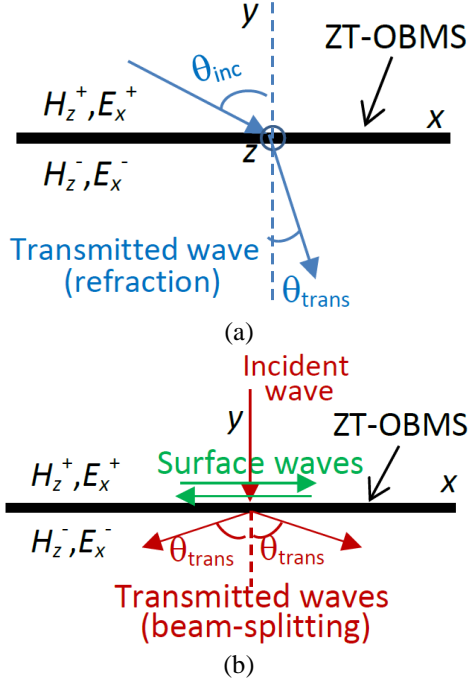

FIG 2. A zero-thickness omega bianisotropic metasurface illuminated by a plane wave. The metasurface with periodic properties produces a discontinuity in the field components that are tangent to the surface. The *H*-field values above and below the surface are $H_z^+$ and $H_z^-$, respectively, and the *x*-components of the *E*-field above and below the surface are $E_x^+$ and $E_x^-$, respectively. (a) Schematic wave configuration for anomalous refraction. (b) Schematic wave configuration for beam-splitting, including non-locally-excited surface waves.

A metasurface is generally designed to produce a prescribed field transformation resulting in known values of $H_z^+(x)$, $H_z^-(x)$, $E_x^+(x)$ and $E_x^-(x)$ along the surface. The design of a metasurface to produce these fields requires determination of $Z_{se}(x)$, $Y_{sm}(x)$ and $K_{em}(x)$. When the power is conserved locally across the surface, Eqs. (1) and (2) lead to [16]

$$K_{em}(x) = \frac{1}{2} \frac{\mathrm{Re}(H_z^- E_x^{+*} - H_z^+ E_x^{-*})}{\mathrm{Re}[(E_x^+ - E_x^-)(H_z^{+*} - H_z^{-*})]}, \qquad (3)$$

$$Z_{se}(x) = i\left[\frac{1}{2}\mathrm{Im}\left(\frac{E_x^+ + E_x^-}{H_z^+ - H_z^-}\right) + K_{em}(x)\,\mathrm{Im}\left(\frac{E_x^+ - E_x^-}{H_z^+ - H_z^-}\right)\right],(4)$$

$$Y_{sm}(x) = i\left[\frac{1}{2}\mathrm{Im}\left(\frac{H_z^+ + H_z^-}{E_x^+ - E_x^-}\right) - K_{em}(x)\,\mathrm{Im}\left(\frac{H_z^+ - H_z^-}{E_x^+ - E_x^-}\right)\right].(5)$$

As indicated previously, the abstract quantities $Z_{se}(x)$, $Y_{sm}(x)$ and $K_{em}(x)$ should be used to obtain the physically realizable transmission coefficient $T(x)$ and reflection coefficient $R(x)$ for normal incidence at each value of *x* along the surface. For any particular value $x=x_p$ of *x*, the $T(x_p)$ and $R(x_p)$ can be found from Eqs. (3) to (5) by assuming local homogeneity of the surface in the region $x=x_p$; that is, by considering the entire surface as characterized by $Z_{se}(x_p)$, $Y_{sm}(x_p)$ and $K_{em}(x_p)$ [10,15,16,23].

To demonstrate this procedure, assume that Eqs. (3) to (5) have been used to find the $Z_{se}(x)$, $Y_{sm}(x)$ and $K_{em}(x)$ which would produce the desired $H(x)$, $E(x)$ field distribution along the surface. It is now desired to find the local transmission and reflection coefficients $T(x_p)$ and $R(x_p)$ at the point $x=x_p$ along the surface. At this point the surface is characterized by $Z_{se}(x_p)$, $Y_{sm}(x_p)$ and $K_{em}(x_p)$ which are known. But it is desired to determine their effect on a wave that is normally incident from above. In this case, the fields above and below the surface coinciding with $y=0$ can be written

$$H_z^>(x,y) = H_0(e^{-iky} + Re^{iky}), \qquad (6)$$

$$E_x^>(x,y) = ZH_0(e^{-iky} - Re^{iky}), \qquad (7)$$

$$H_z^<(x,y) = TH_0 e^{-iky}, \qquad (8)$$

$$E_x^<(x,y) = ZTH_0 e^{-iky}, \qquad (9)$$

where the superscripts > and < indicate the regions $y>0$ and $y<0$, $H_0$ is the amplitude of the incident wave, $Z$ is the impedance of free space, $k=2\pi/\lambda$ is the wave number in free space, $\lambda$ is the wavelength in free space, an $e^{-i\omega t}$ time dependence is suppressed, and $T=T(x_p)$ and $R=R(x_p)$ are the uniform transmission and reflection coefficients of the assumed homogeneous surface with properties $Z_{se}(x_p)$, $Y_{sm}(x_p)$ and $K_{em}(x_p)$. Since

$$H_z^+ = H_z^>(x,0), \quad H_z^- = H_z^<(x,0), \qquad (10)$$

$$E_z^+ = E_z^>(x,0), \quad E_z^- = E_z^<(x,0), \qquad (11)$$

Eqs. (6) to (9) may be evaluated at $y=0$ and used in Eqs. (3) to (5) to yield two linear equations in the two unknowns $T$ and $R$ which may be solved at each point $x=x_p$ as:

$$R = \frac{-4K_{em} + 2\left[-2K_{em}ZY_{sm} + \dfrac{Z_{se}}{Z}\right] + \dfrac{4Z_{se}}{Z}ZY_{sm}}{4K_{em}^2 + 1 + 4Z_{se}Y_{sm} + 2\left(\dfrac{Z_{se}}{Z} + ZY_{sm}\right)} \qquad (12)$$

$$T = \frac{4K_{em}^2 - 1 + 4ZY_{sm}\dfrac{Z_{se}}{Z}}{4K_{em}^2 + 1 + 4Z_{se}Y_{sm} + 2\left(\dfrac{Z_{se}}{Z} + ZY_{sm}\right)} \qquad (13)$$

Since this can be performed for all desired values of $x_p$, it provides the desired solution $T(x)$ and $R(x)$ for all values of $x$. It is these values of $T(x)$ and $R(x)$ that will be mimicked in a structure that produces the same distinctive wave features as those produced by the $Z_{se}(x)$, $Y_{sm}(x)$ and $K_{em}(x)$. To appreciate the connection between the $T(x)$ and $R(x)$ on the one hand and the $Z_{se}(x)$, $Y_{sm}(x)$ and $K_{em}(x)$ on the other hand, it is important to realize that for a *passive and lossless* network, the $Z_{se}$ and $Y_{sm}$ are imaginary while $K_{em}$ is real [16]. The $Z_{se}$, $Y_{sm}$ and $K_{em}$ therefore define three independent parameters which are the three degrees of freedom for each meta-atom. The $T(x)$ and $R(x)$ also represent three independent parameters: the magnitude of $T$, the phase of $T$ and the phase of $R$. The magnitude of $R$ is not independent since

$$|T|^2 + |R|^2 = 1. \qquad (14)$$

In Eqs. (12) and (13), these $T$ and $R$ parameters inherit the three degrees of freedom found in the $Z_{se}(x)$, $Y_{sm}(x)$ and $K_{em}(x)$.

It should therefore not come as a surprise that the same $T$ and $R$ which define (downward) propagation normal to the surface from the incidence region to the transmission region also completely define upward reflection and transmission. This may be understood by realizing based on physical arguments that the transmission $T$ would remain completely unchanged under a change of propagation direction (reciprocity), while the *magnitude* |$R$| of the reflection coefficient would remain unchanged as well (no gain or loss). Since changing the propagation direction is tantamount to changing the sign of $K_{em}$, it can be shown that for a passive lossless surface characterized by $T$ and $R$ for downward propagation, the reflection coefficient for upward propagation may be written

$$Q(x) = -R^*(x) e^{2i\phi_t(x)}, \qquad (15)$$

where $R^*$ denotes the complex conjugate of $R$, and $\phi_t(x)$ is the phase of $T(x)$. As indicated previously, this asymmetry in the reflection coefficient – that is, $R(x) \neq Q(x)$ – is expected for OBMS [14,16,17].

### B. FP-OBMS Design

The design of a real finite thickness structure will now be described that will provide the same $R(x)$, $T(x)$ as the abstract zero-thickness metasurface for any desired application. In Sections IV and V, these principles will be specifically applied to the periodic designs that were chosen herein to demonstrate the wave control capabilities of the structure: anomalous refraction and beam-splitting. As indicated previously, the periodic structure that will be synthesized is composed of an array of meta-atoms, each meta-atom consisting of a narrow parallel plate waveguide containing a cascade of two dielectric layers with the same relative permittivity $\varepsilon_r$ (see Fig. 1) [30]. The width $w_i$ (Fig. 1(b)) of each air-layer and each dielectric-layer is adjusted in a Fabry-Perot manner to provide the desired magnitude and phase of the transmission and reflection coefficients $T$ and $R$.

Assume that the width of each meta-atom is $\Delta \ll \lambda$, and that there are $N_{wg}$ meta-atoms in each period $d = N_{wg}\Delta$ of the structure. The $x$-center of each meta-atom relative to the start of the period will then be located at $x_p = \left(p - \dfrac{1}{2}\right)\Delta, p = 1, 2, ..., N_{wg}$. Since the meta-atom waveguide is narrow, only its fundamental mode will propagate, so that the propagation within the meta-atom is directed in the axial $y$-direction. For the assumed TM propagation the $H$-field is parallel to the PEC walls of the waveguide and the $E$-field is normal to them, so that the fields propagating within each waveguide will be identical to the fields that would be obtained if that waveguide were infinitely thick (i.e. $\Delta \to \infty$). This represents an important advantage of TM polarization. The problem of propagation through the waveguide layers is therefore identical to the one-dimensional problem of propagation of a normally incident wave through homogeneous layers of infinite extent, the well-known solution to which may be written in the form [30,35]

$$T = T(w_1, w_2, w_3, w_4), \quad R = R(w_1, w_2, w_3, w_4). \qquad (16)$$

Note that Eq. (16) does not contain dependence on the lowest air level $w_5$, since that level affects neither $T$ nor $R$. That level is nevertheless required to assure that the lower boundary of each meta-atom coincides with the bottom $y = -h$ of the structure (see Fig. 1). (It will be seen later that $h$ must be an integer multiple of the free space wavelength.)

In order for the meta-atom to emulate the abstract metasurface, its 5-layer structure must provide the same transmission and reflection coefficients as prescribed in Eqs. (12) and (13). That is, for the meta-atom located at $x = x_p$, the layer widths $w_{p1}$, $w_{p2}$, $w_{p3}$, $w_{p4}$ must be found which produce $T = T(x_p) = T_p$, $R = R(x_p) = R_p$, where $w_{pi}$ denotes the $i$th layer in the $p$th meta-atom. This would require the solution of

$$T(w_{p1}, w_{p2}, w_{p3}, w_{p4}) = T_p, \ 1 \le p \le N_{wg},$$
$$R(w_{p1}, w_{p2}, w_{p3}, w_{p4}) = R_p, \ 1 \le p \le N_{wg}. \qquad (17)$$

Eqs. (17) are complex so that they represent four equations (real and imaginary parts of the $T$-equation, and real and imaginary parts of the $R$-equation) in the four unknown layer widths. However, since the magnitudes of $T$ and $R$ are also related through Eq. (14), it is possible to choose only three of the four equations in Eqs. (17), and to solve for three of the four layer widths, with the additional layer width chosen arbitrarily. For example, if $w_{p4}$ is chosen arbitrarily, the system which must be solved may be written

$$\text{Re}[T(w_{p1},w_{p2},w_{p3}) = T_p], \text{Im}[T(w_{p1},w_{p2},w_{p3}) = T_p],$$
$$\text{Re}[R(w_{p1},w_{p2},w_{p3}) = R_p]. \quad (18)$$

As indicated previously, the fact that the proposed realistic metasurface structure requires the solution of three unknowns is entirely compatible with the three degrees of freedom manifested by $Z_{se}$, $Y_{sm}$ and $K_{em}$ in the OBMS sheet transition conditions of Eqs. (1) and (2) [16]. Since the forward problem of Eq. (16) for finding the scattering coefficients of a given unit cell configuration can be readily formulated analytically [30,35], it is clear that the solution to the inverse problem of Eq. (18) may be found with the aid of any standard algorithm for solving three non-linear equations in three unknowns.

Since $w_{p1}$ affects only the phase of $R$, an equivalent possibility that is tactically simpler is to also choose $w_{p1}$ arbitrarily, say $w_{p1}=0$, and to solve *only* the two $T$-equations in Eqs. (18) for $w_{p2}$ and $w_{p3}$:

$$\text{Re}[T(w_{p2},w_{p3}) = T_p], \text{Im}[T(w_{p2},w_{p3}) = T_p], \quad (19)$$

After this is accomplished, the initial air region thickness $w_{p1}$ (which does not affect the value of $T_p$) can be adjusted to provide the desired phase of $R_p$, thereby completing the determination of the three unknown layer thicknesses. That is, if $R_p'$ is the reflection coefficient obtained after solving Eqs. (19) with $w_{p1}=0$, then the required $R_p$ will be obtained by setting

$$w_{p1} = \lambda(\angle R_p - \angle R_p')/4\pi. \quad (20)$$

This approach has the advantage of requiring solution of only two non-linear equations in two unknowns (Eqs. (19)). In any case, it should be clear that the solution for the $w_{pi}$ in Eqs. (18) or (19) is far from unique. Once the widths $w_{pi}$ are set via this procedure for each meta-atom $1 \leq p \leq N_{wg}$, the complete FP-OBMS can be constructed, as shown in Fig. 1(a).

## III. ANALYSIS

In the previous section, an abstract zero-thickness metasurface that can perform some special function (*e.g.* anomalous refraction, beam splitting) was characterized by its local normal-incidence transmission and reflection coefficients $T(x)$ and $R(x)$ as determined by solutions to Eqs. (12) and (13), and a method was presented for designing an electrically thick FP-OBMS with the same transmission and reflection properties. It will be shown in Sections IV and V using full-wave solution methods that such an FP-OBMS does indeed produce the special functionality for which it was designed. Such full-wave solutions are performed for a particular incident wave direction; if this direction is changed, or if any other parameter of the problem is changed, an entirely new and often time-consuming computation would be required. In this section, an analytical method will be derived for rapidly computing scattered fields without the need to consider the intricacies of the FP-OBMS layered configuration. Instead, the $T(x)$ and $R(x)$ conditions, on which this layered configuration is based, are applied to the array of parallel plate waveguides. This analytical tool will be seen not only to provide insight into the operation of the FP-OBMS under non-design conditions, but also to assist in understanding the underlying mechanism of "perfect" anomalous refraction by the OBMS.

Unlike the HMS analytical method used in Ref. [30] for which the local transmission at each $x$-location was perfect ($|T|=1$, $R=0$), the FP-OBMS analytical method developed here will require that $R(x)$ be finite, and will depend on the value of $x$ along the metasurface such as the one shown in Fig. 1(a). Indeed, it is this non-perfect transmission which leads to the advantages of the OBMS [14,16], and must be accounted for when describing the propagation through each waveguide of the array.

Each waveguide is assumed sufficiently narrow to allow only the fundamental mode to propagate within it. If waveguide $p$ were empty, the field within it could be written as the sum of an upward wave and a downward wave: $H_{2p} = \sigma_{pa}e^{-iky} + \sigma_{pb}e^{iky}$, where $\sigma_{pa}$ and $\sigma_{pb}$ are the respective wave amplitudes. Although it is *not* empty, the transmission and reflection coefficients $T_p$, $R_p$ and $Q_p$ are known from Eqs. (12), (13) and (15), and are sufficient for formulating boundary conditions which only require the field values at the waveguide apertures. The field within waveguide $p$ may therefore be written using separate expressions near its top and near its bottom:

$$H_{2p}(y) = \begin{cases} \sigma_{pa}e^{-iky} + (T_p\sigma_{pb} + R_p\sigma_{pa})e^{iky}, & y \to 0^-, \\ (T_p\sigma_{pa} + Q_p\sigma_{pb})e^{-iky} + \sigma_{pb}e^{iky}, & y \to -h^+, \end{cases} \quad (21)$$

where $1 \leq p \leq N_{wg}$, and the amplitudes $\sigma_{pa}$ and $\sigma_{pb}$ are to be determined by enforcing the boundary conditions along the facets of the entire metasurface. Eq. (21) will, of course, reduce to the analogous HMS equation (19) of [30] when $R_p \to 0$. $H_{2p}(y)$ depends on $x$ through $p$ which is a discrete function of $x$: $p=1+\text{int}(x/\Delta)$.

To understand the right side of Eq. (21), consider the field near the top of the waveguide ($y \to 0^-$). The "incident" downward wave there, $\sigma_{pa}e^{-iky}$, will have the same form as that for an empty waveguide. The upward wave there started as an "incident" wave $\sigma_{pb}e^{iky}$ at the bottom, and reached the top after propagating through the dielectric loading for which the transmission coefficient is $T_p$. But there is an additional component to this field: the portion of the incident wave $\sigma_{pa}e^{-iky}$ that is reflected with reflection coefficient $R_p$. The amplitude of the upward wave at the top is therefore $T_p\sigma_{pb}e^{iky} + R_p\sigma_{pa}e^{iky}$, as given in the $y \to 0^-$ expression in Eq. (21). The $y \to -h^+$ expression can be understood in the same manner except now the downward reflection coefficient $R_p$ is replaced by the upward reflection coefficient $Q_p$.

Details of the analytical derivation for the scattered fields are given in the Appendix. It is based on Floquet-Bloch (FB) expansions of the $H$-field in the incidence-reflection region $y>0$ (where the modal reflection amplitudes are denoted $\rho_n$), and in the transmission region $y<-h$ (where the modal transmission amplitudes are denoted $\tau_n$) [30]. Within each waveguide the field is given by Eq. (21). Continuity of the $H$ and $E$ components parallel to the FP-OBMS across its upper surface at $y=0$ and across its lower surface at $y=-h$ provide four equations in the unknown amplitudes. The waveguide amplitudes $\sigma_{pa}$, $\sigma_{pb}$ can be eliminated from these four equations, leaving two Fourier series in the unknown amplitudes $\rho_n$, $\tau_n$ which depend on:
(a) the design parameters $\theta_{inc}$, $\theta_{trans}$;
(b) $R(x)$, $T(x)$, $Q(x)$, which may themselves be written as Fourier series,

$$R(x) = \sum_{s=-\infty}^{\infty} a_{rs} e^{i\left(\frac{2s\pi}{|d|}\right)x}, \quad T(x) = \sum_{s=-\infty}^{\infty} a_{ts} e^{i\left(\frac{2s\pi}{|d|}\right)x},$$

$$Q(x) = \sum_{s=-\infty}^{\infty} a_{qs} e^{i\left(\frac{2s\pi}{|d|}\right)x}, \quad (22)$$

$$d = \frac{\lambda}{\Delta_{\sin}}, \quad \Delta_{\sin} = \sin\theta_{trans} - \sin\theta_{inc}, \quad (23)$$

$|d|$ is the period; and
(c) the actual incidence angle $\psi_{inc}$ which may be different from the design incidence angle $\theta_{inc}$ for which the FP-OBMS was constructed.

The procedure in the Appendix results in an infinite number of linear equations (Eqs. (A25) and (A26)) in an infinite number of unknowns, which may be truncated to

$$-\left(\rho_m C_m - \sum_{s=-K}^{K} a_{r(m-s)} \rho_s S_s\right) + e^{ikh} \sum_{s=-K}^{K} a_{t(m-s)} \tau_s S_s,$$
$$= \delta_{m0} S_0 - a_{rm} C_0, \quad -K \leq m \leq K, \quad (24)$$

$$-\sum_{s=-K}^{K} a_{t(m-s)} \rho_s S_s + e^{-ikh}\left(\tau_m C_m - e^{2ikh}\sum_{s=-K}^{K} a_{q(m-s)} \tau_s S_s\right),$$
$$= a_{tm} C_0, \quad -K \leq m \leq K, \quad (25)$$

where $K$ is a positive integer, and the Fourier series coefficients $a_{ts}$, $a_{rs}$, $a_{qs}$ are defined from Eqs. (22). The $S_s$ and $C_s$ are defined as

$$S_s = \frac{1}{2}(1-\cos\psi_s), \quad C_s = \frac{1}{2}(1+\cos\psi_s), \quad (26)$$

$$\cos\psi_s = \sqrt{1-\alpha_s^2/k^2}, \quad \alpha_s = k\sin\psi_{inc} + \frac{2s\pi}{d}, \quad (27)$$

and are related to the Fresnel reflection coefficient $\Gamma_s$. For a plane wave incident on a surface at an angle $\psi_s$ that is refracted into a plane wave at angle $\theta_{trans}$, the Fresnel reflection coefficient would be $\Gamma_s$ for transverse electric (TE) waves considered in [12], and $-\Gamma_s$ for the TM waves being considered here, where

$$\Gamma_s(\psi_s, \theta_{trans}) = \frac{\cos\theta_{trans} - \cos\psi_s}{\cos\theta_{trans} + \cos\psi_s}. \quad (28)$$

Interestingly, from (26) it turns out that (28) with $\theta_{trans}=0$ leads to,

$$\frac{S_s}{C_s} = \frac{1-\cos\psi_s}{1+\cos\psi_s} = \Gamma_s(\psi_s, 0). \quad (29)$$

It will be recalled that our goal is to utilize the FP-OBMS to simulate the abstract zero-thickness OBMS characterized by $h\to 0$, or $e^{ikh}=1$. Since Eqs. (24) and (25) – which describe the fields scattered from the FP-OBMS – contain both $e^{ikh}$ and $e^{-ikh}$ factors, values of $h$ other than $n\lambda$ will produce solutions $\rho_m$, $\tau_m$ with magnitudes that are different from those of interest. Therefore, in what follows, values of $h$ will be restricted to those for which $e^{ikh}=1$. Such values can be attained by adjusting the thickness of the lowest air-layer of the FP-OBMS.

Eqs. (24) and (25) represent $2(2K+1)$ simultaneous linear equations in the same total number of unknown FB field amplitudes $\rho_m$ and $\tau_m$. These fields may be solved for a wave incident at any angle $\psi_{inc}$ on the FP-OBMS that has been designed to provide the scattered fields of interest. This design is characterized by the periodic surface susceptibilities $Z_{se}$, $Y_{sm}$ and $K_{em}$ which have been translated to local reflection and transmission coefficients $R$, $T$ and $Q$. The solution of Eqs. (24) and (25) is straightforward, avoiding the need for macroscopic full-wave simulations in commercial solvers.

In what follows, the above developments will be applied to FP-OBMS structures that are designed for anomalous refraction, and for beam-splitting. For each of these applications, the ability of the thick structure to perform the desired function will be demonstrated. In addition, the underlying mechanism employed by the OBMS to produce anomalous refraction will be described.

## IV. ANOMALOUS REFRACTION

For the anomalous refraction application, it is desired that all the energy of a plane wave incident on the OBMS at an angle $\theta_{inc}$ be transmitted into a plane wave at an angle $\theta_{trans}$ (see Fig. 2(a)). It will now be shown that the FP-OBMS structure can successfully fulfill this requirement. It will further be shown that solutions of Eqs. (24) and (25) successfully predict the fields scattered by the anomalous refraction metasurface, both for incident fields in the design direction $\theta_{inc}$ and for incident fields in other directions $\psi_{inc}$.

For the specific case of $\theta_{trans}=0$, these reduce to closed form solutions which can be employed to compare OBMS fields with HMS fields. Such a comparison will be shown to reveal that the underlying mechanism of OBMS superiority over the HMS lies in its ability to provide a virtual anti-reflective coating for the metasurface.

### A. FP-OBMS Structure for Anomalous Refraction

In order to define the structural layers with the aid of Eqs. (18) or (19), or to utilize Eqs. (24) and (25) which govern the fields scattered from the FP-OBMS, the reflection and transmission coefficients $R(x)$ and $T(x)$ for normal incidence must be known. The scattered fields for this case along a zero-thickness metasurface will satisfy [14,16,17]

$$H_z^>(x,y) = H_0 e^{ikx\sin\theta_{inc}} e^{-iky\cos\theta_{inc}} \quad (30)$$

$$E_x^>(x,y) = H_0 Z \cos\theta_{inc} e^{ikx\sin\theta_{inc}} e^{-iky\cos\theta_{inc}} \quad (31)$$

$$H_z^<(x,y) = H_0 \sqrt{\frac{\cos\theta_{inc}}{\cos\theta_{trans}}} e^{ikx\sin\theta_{trans}} e^{-iky\cos\theta_{trans}} \quad (32)$$

$$E_x^<(x,y) = H_0 Z \cos\theta_{trans} \sqrt{\frac{\cos\theta_{inc}}{\cos\theta_{trans}}} e^{ikx\sin\theta_{trans}} e^{-iky\cos\theta_{trans}} \quad (33)$$

The square root factor is present in Eqs. (32) and (33) to assure that power flow across the surface is conserved. Using Eqs. (30) to (33) in Eqs. (10) and (11), and the result in Eqs. (3) to (5) will yield the values of $K_{em}(x)$, $Z_{se}(x)$ and $Y_{sm}(x)$. Using these, in turn, in Eqs. (12) and (13) will enable $T(x)$ and $R(x)$ to be found as

$$T(x) = \frac{e^{i2\pi x/d}\sqrt{\cos\theta_{inc}\cos\theta_{trans}}}{\left(\cos\frac{\theta_{inc}}{2}\cos\frac{\theta_{trans}}{2}\right)^2 - e^{i4\pi x/d}\left(\sin\frac{\theta_{inc}}{2}\sin\frac{\theta_{trans}}{2}\right)^2}, \quad (34)$$

$$R(x) = \frac{\left(\sin\frac{\theta_{inc}}{2}\cos\frac{\theta_{trans}}{2}\right)^2 - e^{i4\pi x/d}\left(\cos\frac{\theta_{inc}}{2}\sin\frac{\theta_{trans}}{2}\right)^2}{\left(\cos\frac{\theta_{inc}}{2}\cos\frac{\theta_{trans}}{2}\right)^2 - e^{i4\pi x/d}\left(\sin\frac{\theta_{inc}}{2}\sin\frac{\theta_{trans}}{2}\right)^2}. \quad (35)$$

For $\theta_{inc}=80°$, Fig. 3 illustrates the functions $R(x)$ and $T(x)$ for two different refraction scenarios, $\theta_{trans}=0°$ (Fig. 3(a)) and $\theta_{trans}=30°$ (Fig. 3(b)).

The $R(x)$ and $T(x)$ as obtained from Eqs. (34) and (35) may be used in Eqs. (18) or (19) to obtain the $w_i$ for each waveguide element of the FP-OBMS at $x=x_p$ along the metasurface. In this manner, the $R(x)$ and $T(x)$ of Fig. 3(a) leads to the structure shown in Fig. 1(a) for the case $\theta_{inc}=80°$, $\theta_{trans}=0$; and the $R(x)$ and $T(x)$ of Fig. 3(b) leads to the structure shown in Fig. 4 for the case $\theta_{inc}=80°$, $\theta_{trans}=30°$. For these anomalous refraction FP-OBMS structures in Figs. 1 and 4, the asymmetry relative to their centers at $y=-h/2$ is again consistent with the known association of abstract OBMSs with an *asymmetric* cascade of three electric-impedance sheets [14,16-18]. This is in contrast to the analogous FP-HMS which was symmetric about $y=-h/2$ [30], consistent with three *symmetric* electric-impedance sheets for the zero-thickness abstract case [6, 33,36].

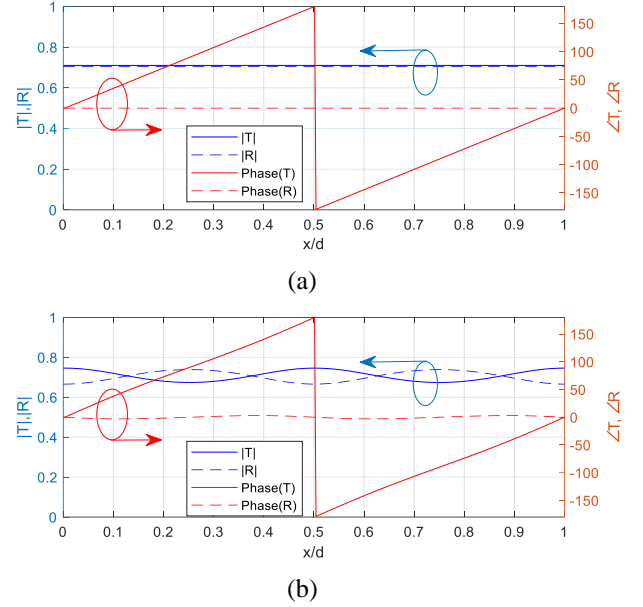

(a)

(b)

FIG. 3. The magnitudes and phases of $T(x)$ (solid curves) and $R(x)$ (dashed curves) of Eqs. (12) and (13) for anomalous refraction. (a) $\theta_{inc}=80°$, $\theta_{trans}=0°$, (b) $\theta_{inc}=80°$, $\theta_{trans}=30°$. The magnitudes are referenced to the left-hand axis, and the phases are referenced to the right-hand axis.

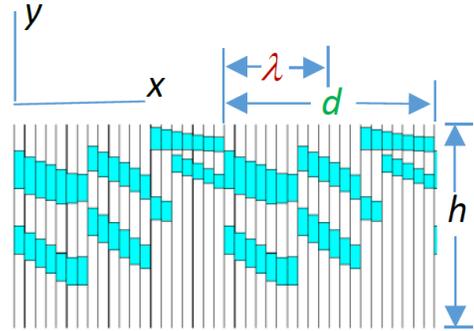

FIG. 4. Two periods of the FP-OBMS obtained from Eqs. (19) for anomalous refraction, $\theta_{inc}=80°$, $\theta_{trans}=30°$, $h=2.0\lambda$, $d=2.0627\lambda$, $N_{wg}=20$. The colored region is the dielectric $\varepsilon_r=16$, the white region is air. This FP-OBMS produces the $T(x)$ and $R(x)$ of Fig. 3(b).

### B. Design Validation

The spatial fields scattered from the FP-OBMS structure in Fig. 4 are illustrated in Fig. 5 as calculated by both the full-wave CST solution [37] and by the analytical solution to Eqs. (24) and (25). In both cases, an incident wave in the design direction $\psi_{inc}=\theta_{inc}=80°$ was assumed. From the CST results, 99.8% of the incident energy was transferred to the anomalously refracted wave for which $\psi_{trans}=30°$, indicating that the FP-OBMS achieves perfect anomalous for this wide-refraction-angle scenario. In contrast, for an HMS, over 44% of the energy would be lost to spurious reflections. Note that Fig. 5 displays the field both within and outside the FP-OBMS. The good agreement in Fig. 5 between the CST full-wave and the analytical solutions, both within and outside the FP-OBMS, verifies the fidelity of the analytical model calculations.

Fig. 5 provides results for a wave incident on the structure of Fig. 4 in the design direction $\theta_{inc}$. Solutions to Eqs. (24) and (25) may also be obtained for waves incident on the same structure from other directions $\psi_{inc} \neq \theta_{inc}$. In Fig. 6, power coupling efficiencies are provided as functions of $\psi_{inc}$. The power coupling efficiency to the anomalous refraction mode is given by $\eta_{-1}^\tau = |\tau_{-1}|^2 \cos\psi_{trans}/\cos\psi_{inc}$, while the power coupling efficiency to the specular reflection mode is $\eta_0^\rho = |\rho_0|^2$. These are presented in Fig. 6 using the analytical solution to Eqs. (24) and (25), and using full-wave results obtained by applying CST to the FP-OBMS structure shown in Fig. 4. The agreement in Fig. 6 between these two computational methods verifies the usefulness of the analytical model as an alternative to full-wave solutions even for determining the angular response of FP-OBMSs.

### C. OBMS as HMS with Virtual Anti-Reflective Coating

It will now be shown that the underlying mechanism of OBMS superiority over the HMS derives from its ability to form a virtual anti-reflective coating (ARC) for the metasurface [38,39]. This will be accomplished by comparing an anomalous refraction solution of Eqs. (24) and (25) for the OBMS with the previously found analogous solution for the HMS [30]. These solutions will then be shown to be consistent with geometrical optics models of the surfaces which are identical except for a virtual ARC that is present on the OBMS but absent on the HMS.

#### 1. Closed-Form Analytics

For analytical simplicity, normally directed transmission $\theta_{trans}=0°$ will be assumed, whereas this restriction will be relaxed later. Then Eqs. (34), (35) and (15) reduce to

$$T(x) = \sqrt{1-r^2}\, e^{i2\pi x/d}, \quad R(x) = r, \quad Q(x) = -r e^{i4\pi x/d}, \quad \theta_{trans}=0, \tag{36}$$

where, in analogy to Eq. (28),

$$r(\theta_{inc}, \theta_{trans}) = \frac{\cos\theta_{trans} - \cos\theta_{inc}}{\cos\theta_{trans} + \cos\theta_{inc}} \underset{\theta_{trans} \to 0}{=} \tan^2\frac{\theta_{inc}}{2}. \tag{37}$$

From Eq. (36), it is immediately apparent that the Fourier coefficients in Eqs. (22) are $a_{ts} = \sqrt{1-r^2}\,\delta_{s,-1}$, $a_{rs}=r\delta_{s,0}$, $a_{qs}=-r\delta_{s,-2}$ so that, for the OBMS, Eqs. (24) and (25) reduce to

$$-\left(C_m - rS_m\right)\rho_m + \sqrt{1-r^2}\,\tau_{m+1} S_{m+1} = \delta_{m0}(S_0 - rC_0), -K \leq m \leq K, \tag{38}$$

$$-\sqrt{1-r^2}\,\rho_{m+1} S_{m+1} + \left(\tau_m C_m + r\tau_{m+2} S_{m+2}\right) = \sqrt{1-r^2}\,\delta_{m,-1} C_0, -K \leq m \leq K, \tag{39}$$

where, as previously discussed, we have set $e^{ikh}=1$.

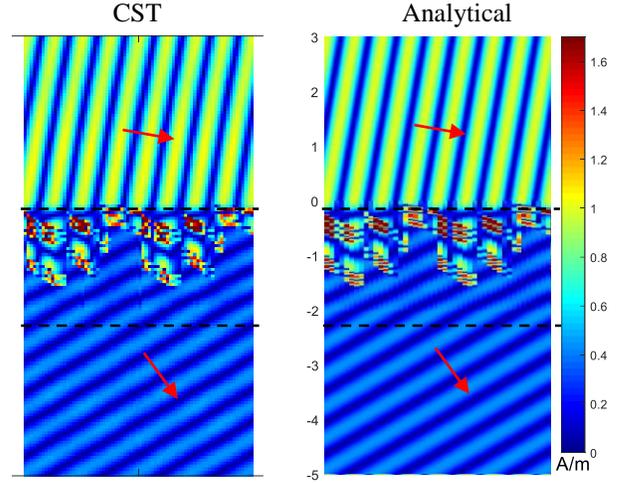

FIG. 5. H-field color images of propagation through the structure of Fig. 4 which was designed to anomalously refract an incident wave at $\theta_{inc}=80°$ into a transmitted wave at $\theta_{trans}=30°$. Displayed are full-wave computations performed by CST, and the analytical solution of Eqs. (24) and (25) for actual wave incidence angle $\psi_{inc}$ equal to the design angle of incidence. All geometric scales are relative to a wavelength. The horizontal extent of each plot is two periods.

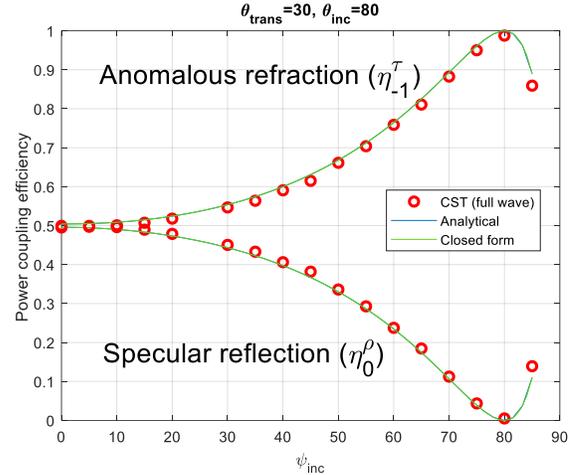

FIG. 6. Power coupling efficiency comparisons between CST full-wave solutions applied to the FP-OBMS shown in Fig. 4, the analytical solution of Eqs. (24) and (25), and the closed form solution of Eq. (42) (see Section IV.C). The latter two curves are essentially coincident.

These simultaneous linear equations may be solved for the Floquet-Bloch (FB) field amplitudes $\rho_m$, $\tau_m$ for any actual incidence angle $\psi_{inc}$, where the dependence on $\psi_{inc}$ enters through $S_m$, $C_m$ (see Eq. (26)). But since our primary goal is to compare the OBMS with the HMS for operation at the angles for which they were designed, we will scrutinize the particular case for which $\psi_{inc}$ equals the design angle of incidence $\theta_{inc}$. In this case, $\psi_{trans}=\theta_{trans}=0$, and from Eq. (A20) $C_{-1}=1$, $S_{-1}=0$. Then, for the OBMS, Eqs. (38) and (39) are satisfied for $\rho_m=\tau_m=0$ for all $m$ except, by applying Eq. (38) for $m=0$ and Eq. (39) for $m=-1$:

$$\rho_0 = -\frac{S_0 - rC_0}{C_0 - rS_0} = -\frac{\Gamma_0 - r}{1 - r\Gamma_0}, \quad (40)$$

$$\tau_{-1} = \sqrt{1-r^2}\,\frac{C_0 + S_0\rho_0}{C_{-1}} = \frac{\sqrt{1-r^2}(1-\Gamma_0)(1+\Gamma_{-1})}{1 - r\Gamma_0}, \quad (41)$$

where the $\Gamma_n=\Gamma_n(\theta_{inc},0)$ were defined in Eq. (28) and $r=r(\theta_{inc},0)$ was defined in Eq. (37). From Eq. (A19) and the assumed condition $\psi_{inc}=\theta_{inc}$, it follows that $\Gamma_0=r$ so that, from Eq. (40), $\rho_0$ also vanishes, and $\tau_{-1}$ is the lone surviving Floquet-Bloch amplitude as originally sought. In addition, $\tau_{-1}$ is precisely the target field given in Eq. (32). In spite of the fact that $\rho_0$ vanishes as does $\Gamma_{-1}$, Eqs. (40) and (41) will be kept as shown in order to later facilitate their application to more general values of $\psi_{inc}$, $\theta_{inc}$ and $\theta_{trans}$. Thus, to summarize, the design fields for the OBMS are

$$\rho_0 = -\frac{\Gamma_0 - r}{1 - r\Gamma_0},\ \tau_{-1} = \frac{\sqrt{1-r^2}(1-\Gamma_0)(1+\Gamma_{-1})}{1 - r\Gamma_0},\ \text{OBMS}. \quad (42)$$

We previously derived the analogous expressions for the FP-HMS which are given by [30]:

$$\rho_0 = -\Gamma_0, \quad \tau_{-1} = (1-\Gamma_0)(1+\Gamma_{-1}), \text{ HMS}. \quad (43)$$

It is worth noting again that Eqs. (42) and (43) were derived for the case in which the wave exits normally from the surface, leading to $\Gamma_{-1}=0$. For the geometrical optics (GO) models which will be described now, it will be sufficient to use $\Gamma_{-1}\ll 1$.

## 2. Geometrical Optics Models

We will now give a geometrical optics (GO) interpretation to the wave propagation across the metasurface, similar to that utilized for HMS in [12]. Although [12] deals with all Floquet-Bloch modes within the HMS, the discussion here will be limited to waves that affect only the specular reflection component $\rho_0$ and the anomalous refraction transmission component $\tau_{-1}$ of both the HMS and the OBMS, which are typically the dominant ones. In this GO interpretation, the zero-thickness metasurface is pictured in Fig. 7 as a virtual region with thickness, but this thickness does not cause phase accumulation of a wave passing through it. The virtual finite thickness, however, does allow for consideration of multiple reflections between the region facets denoted $V_{top}$ and $V_{bot}$. As discussed following Eq. (25), the reflection coefficient $R_n$ of a single ray incident on an interface at angle $\psi_n$ is given by $-\Gamma_n$. This, together with the transmission coefficient $T_n$ are

$$R_n = -\Gamma_n,\ T_n = 1 - \Gamma_n. \quad (44)$$

Since each such reflection of a ray can produce an additional contribution to the total specular reflection mode amplitude $\rho_0$ and to the total anomalous refraction mode amplitude $\tau_{-1}$, it is convenient to write these amplitudes as sums of these ray contributions, similar to a standard dielectric slab [40]:

$$\rho_0 = \rho_0^{(0)} + \rho_0^{(1)} + \rho_0^{(2)} + \cdots, \quad (45)$$

$$\tau_{-1} = \tau_{-1}^{(0)} + \tau_{-1}^{(1)} + \tau_{-1}^{(2)} + \cdots. \quad (46)$$

In the virtual region, the direction of the plane wave is $\theta_{trans}$ [12]. The region above $V_{top}$ contains the incident and reflected waves, while the region below $V_{bot}$ contains the transmitted waves. In order to contrast the HMS with OBMS, the HMS is shown in Fig. 7(a) and the OBMS is shown in Fig. 7(b).

Consider first the HMS case [7,12] in Fig. 7(a). The incident wave is reflected from $V_{top}$, so that from Eq. (44) $\rho_0^{(0)}=-\Gamma_0$. But $\rho_0$ can also have contributions from components that are reflected one or more times from $V_{bot}$. For example, for the contribution $\rho_0^{(1)}$ from only a single reflection from $V_{bot}$, the wave magnitude would be the product of (a) the transmission coefficient $(1-\Gamma_0)$ across $V_{top}$ from the upper region to the virtual region; (b) the reflection coefficient $\Gamma_{-1}$ from $V_{bot}$; and (c) the transmission coefficient $(1+\Gamma_0)$ across $V_{top}$ from the virtual region into the incidence region:

$$\rho_0^{(1)} = (1-\Gamma_0)\Gamma_{-1}(1+\Gamma_0) \quad (47)$$

The contribution after two reflections would be the same as Eq. (47) but with additional factors indicating reflection $\Gamma_0$ from $V_{top}$ and reflection $\Gamma_{-1}$ from $V_{bot}$:

$$\rho_0^{(2)} = (1-\Gamma_0)\Gamma_{-1}\Gamma_0\Gamma_{-1}(1+\Gamma_0). \quad (48)$$

In cases in which $\Gamma_{-1}\ll 1$, $\rho_0^{(n)}\approx 0$, $n>0$, implying $\rho_0\approx\rho_0^{(0)}=-\Gamma_0$ in agreement with Eq. (43) [12].

The HMS expression for $\tau_{-1}^{(0)}$ is the product of the transmission coefficient $(1-\Gamma_0)$ across $V_{top}$ from the upper region to the virtual region, and the transmission coefficient $(1+\Gamma_{-1})$ across $V_{bot}$ from the virtual region into the transmission region: $\tau_{-1}^{(0)}=(1-\Gamma_0)(1+\Gamma_{-1})$. Additional contributions from reflections within the virtual region would again include reflections $\Gamma_{-1}$ from $V_{bot}$ which would be small when $\Gamma_{-1} \ll 1$, so that $\tau_{-1} \approx \tau_{-1}^{(0)} = (1-\Gamma_0)(1+\Gamma_{-1})$ which again agrees with Eq. (43). The GO model shown in Fig. 7(a) for the HMS has thus been shown to be consistent with the HMS-scattered fields given in Eq. (43).

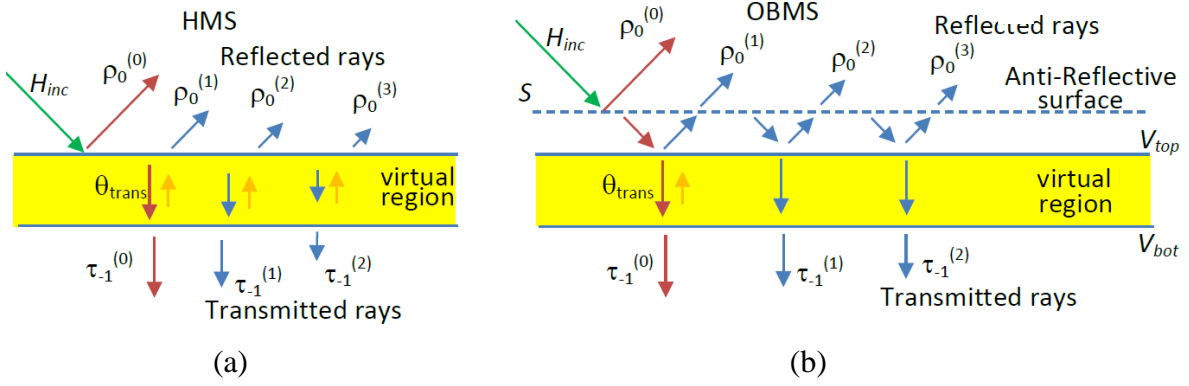

FIG. 7. Zero-thickness metasurfaces for geometrical optics analysis represented by a virtual region through which waves can pass and within which waves can be multiply reflected, but without phase accumulation. (a) An HMS consisting of the virtual region alone. (b) An OBMS consisting of the virtual region adjacent to an anti-reflective surface.

The goal now is to provide a GO model which can produce the OBMS fields in Eq. (42). It will be recalled that although the HMS can be characterized by a symmetric structure, the OBMS will be asymmetric. As shown in Fig. 7(b), this asymmetry for the OBMS will be obtained by adding a virtual structural component $S$ to the top of a virtual region that is identical to that used for the HMS in Fig. 7(a). The properties of this surface $S$ will differ, however, from those of the main "virtual region" surface. For the anomalous refraction application considered here, the virtual region causes the trajectory change of the wave, which lies at the basis of this functionality. The parameters $\Gamma_0$ and $\Gamma_{-1}$ characterize the reflections that occur because of such changes in wave direction (wave-impedance mismatch [10,12]). On the other hand, the surface $S$ acts as a virtual anti-reflective coating which does not affect the direction of the wave [38]. Being a virtual anti-reflective coating, the wave will not change its course across S, but will incur reflection which will destructively interfere with the reflection from the virtual region. More explicitly, an impinging wave will be reflected from S with a reflection coefficient $r$, which is the negative of the reflection coefficient $-\Gamma_0$ from the HMS. Since $r$ as given in Eq. (37) is real, the transmission across $S$ will simply be $(1-r^2)^{1/2}$.

The components of $\rho_0$ and $\tau_{-1}$ can now be constructed for the OBMS in a manner similar to that used above for the HMS. As before, components that are reflected from $V_{bot}$ may be ignored. There will nevertheless be multiple reflections between the virtual anti-reflective surface $S$ and $V_{top}$ that *will* contribute to both $\rho_0$ and $\tau_{-1}$. Referring to Fig. 7(b), for $\rho_0$,

$$\rho_0^{(0)} = r, \tag{49}$$

$$\rho_0^{(1)} = \sqrt{1-r^2}(-\Gamma_0)\sqrt{1-r^2} = -(1-r^2)\Gamma_0, \tag{50}$$

$$\rho_0^{(2)} = -(1-r^2)\Gamma_0(r\Gamma_0), \tag{51}$$

$$\rho_0^{(n)} = -(1-r^2)\Gamma_0(r\Gamma_0)^{n-1}. \tag{52}$$

Therefore, from Eq. (45)

$$\rho_0 = r - (1-r^2)\Gamma_0[1 + r\Gamma_0 + (r\Gamma_0)^2 + (r\Gamma_0)^3 + \cdots] \tag{53}$$

Summing the geometric series in the square brackets leads to

$$\rho_0 = r - \frac{(1-r^2)\Gamma_0}{1-r\Gamma_0} = \frac{r-\Gamma_0}{1-r\Gamma_0}. \tag{54}$$

Similarly, for $\tau_{-1}$,

$$\tau_{-1} = \frac{\sqrt{1-r^2}(1-\Gamma_0)(1+\Gamma_{-1})}{1-\Gamma_0 r}. \tag{55}$$

Since the GO results in Eqs. (54) and (55) correspond precisely to Eq. (42), it may be concluded that **employing an omega bianisotropic metasurface for anomalous refraction is tantamount to adding a virtual anti-reflective coating to a Huygens' metasurface.** This coating produces a reflected wave that is exactly out-of-phase with the unwanted HMS reflected wave (see Eq. (54)), thereby cancelling it entirely. This is in full compliance with symmetry/asymmetry requirements, and with the known properties of the two classes of metasurfaces at the designated angle of incidence. But importantly, the GO model can *also* provide insight into these metasurfaces' scattering properties for other angles of incidence; i.e., it characterizes their angular response. This will now be discussed.

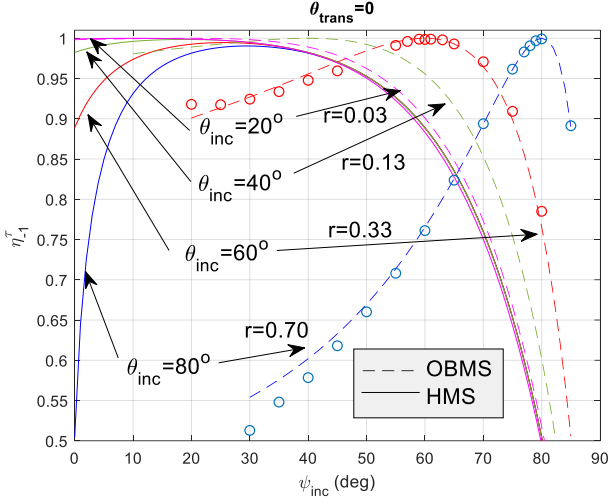

FIG. 8. The HMS and OBMS power coupling efficiency $\eta_{-1}^{\tau}$ as a function of actual incidence angle $\psi_{inc}$ for design transmission angle $\theta_{trans}=0$ and for several values of design incidence angle $\theta_{inc}$. The dashed curves are for OBMS based on Eq. (42) and the solid curves are for HMS based on Eq. (43). The circles represent full wave CST results applied to the $\theta_{inc}=80°$ structure shown in Fig. 1 and a similarly designed structure for $\theta_{inc}=60°$. The HMS $\theta_{inc}=80°$ results had been reported previously and shown to agree well with full wave CST results [30]. The value $r$ of the virtual anti-reflective coating for each OBMS case is shown near the arrow pointing to the respective OBMS results.

### 3. Angular Response

It was shown that the GO models in Fig. 7 are entirely consistent with the field solutions Eqs. (56) and (43) for the OBMS and HMS, respectively. These solutions are plotted in Fig. 8 in terms of the power coupling efficiency $\eta_{-1}^{\tau}$ to the anomalous refraction mode, as a function of the actual (non-design) incidence angle $\psi_{inc}$ for several values of the design incidence angle $\theta_{inc}$. Since the GO derivation of Eq. (55) required $\Gamma_{-1} \ll 1$, Fig. 8 only contains OBMS plots (dashed curves) for values of $\psi_{inc}$ for which this condition is satisfied. The veracity of these closed form solutions is verified by comparing them to full-wave solutions for the larger values of $\theta_{inc}$. The agreement between them wanes at smaller values of $\psi_{inc}$ where the condition $\Gamma_{-1} \ll 1$ is not satisfied.

The OBMS and HMS solutions differ in that the parameter $r$ is present only in the OBMS solution (Eq.(57)) which, as discussed, indicates the presence in the OBMS of the virtual anti-reflective coating. That is, $r$ represents the strength of the virtual anti-reflective coating of Fig. 7(b) that is affixed to the HMS to neutralize its reflection thereby forming the OBMS. This anti-reflective "strength" is defined in Eq. (37) which indicates that $r$ approaches zero rapidly as the design incidence angle $\theta_{inc}$ becomes small (see Fig. 8). Since the purpose of the anti-reflective coating is to neutralize the HMS reflection, and since such reflection is greatest when there is a significant difference in wave impedance between both sides of the surface, it is indeed to be expected that $r$ would be greatest for incident angles $\theta_{inc}$ that differ greatly from the assumed $\theta_{trans}=0$ of the refracted wave. On the other hand, when $\theta_{inc}$ is small, the wave impedance mismatch is small and the HMS would be only mildly reflective.

Indeed, it is clearly seen in Fig. 8 that for $\theta_{inc}=80°$ and $\theta_{inc}=60°$, the OBMS and HMS results are completely different; the OBMS design point at which $\eta_{-1}^{\tau}=1$ is completely isolated, with its value decreasing relatively rapidly at either side of this point. This demonstrates the resonant nature of the virtual anti-reflective coating in these cases of large wave-impedance mismatch, where strong reflection cancellation is necessary (i.e., a resonant "structure"). For these high values of $\theta_{inc}$, the HMS inevitably produces reflection [7,10,13]. Indeed, it is this shortcoming of the HMS which prompted the development of the OBMS. However, for smaller values of $\theta_{inc}$ (resulting in smaller values of $r$), the OBMS results begin to approach those for the HMS. For $\theta_{inc}=40°$, the OBMS results are closer to the HMS results, particularly for $\psi_{inc}<\theta_{inc}$, while for $\theta_{inc}=20°$, the HMS and OBMS are almost coincident, and there is no clear advantage to employing OBMS over HMS.

Since the above development assumed $\Gamma_{-1} \ll 1$, the fields in Eq. (42) would be valid for combinations of $\psi_{inc}$, $\theta_{inc}$, $\theta_{trans}$ satisfying this condition. It may be shown that this condition is met for all values of $\psi_{inc}$ when $\theta_{inc}=80°$, $\theta_{trans}=30°$ which are the parameters employed previously in Fig. 6. In fact, we plotted in Fig. 6 the predictions of the GO model as manifested in Eqs. (54) and (55) and Fig. 7 for this case as well (solid green line). Astoundingly, the intuitive physical model yields results which are indistinguishable from the results presented based on the numerical solution of Eqs. (24) and (25), pointing out its high fidelity and usefulness for a wide range of refracting OBMSs.

## V. BEAM SPLITTING

For the beam splitting application, it is desired to have the OBMS transform a plane wave that is normally incident on it into *two* transmitted plane waves of equal power: one for which the transmission angle is $+\theta_{trans}$, and the other for which the transmission angle is $-\theta_{trans}$ (see Fig. 2(b)). This case requires special processing since the incident plane wave and the two transmitted plane waves do not fulfill the local power conservation requirement along the surface at $y=0$ [16]. However, it has been shown for the beam-splitting case that stipulating two additional auxiliary surface waves propagating on the top facet would facilitate local power conservation [18]. Note that this scheme differs substantially from the typical phase-gradient metasurface design approach, as it requires the metasurface to exhibit a *non-*

*local* response to ensure reflectionless splitting, where the power flow is modulated along the metasurface plane via the excited surface waves.

We utilize this scheme to demonstrate and verify the performance of the FP-OBMS for this beam-splitting application. The fields for this case will satisfy [18]

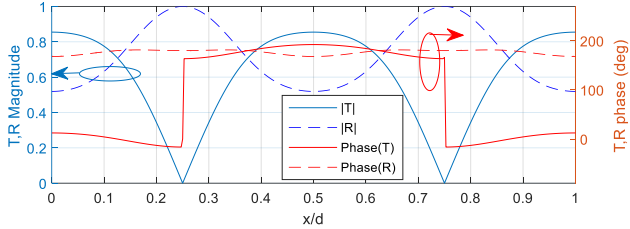

FIG. 9. The magnitudes and phases of $T(x)$ (solid curves) and $R(x)$ (dashed curves) of Eqs. (12) and (13) over a period of a metasurface designed to split an incident wave normal to the surface into two symmetric transmitted waves propagating at angles $\pm\theta_{trans}$, $\theta_{trans}=80°$. The magnitudes are referenced to the left-hand axis, and the phases are referenced to the right-hand axis.

$$H_z^>(x,y) = H_0 e^{ikx\sin\theta_{inc}} e^{-iky\cos\theta_{inc}} + H_0 e^{-a_{sw}y}\cos(2kx\sin\theta_{trans}), \quad (58)$$

$$E_x^>(x,y) = Z\cos\theta_{inc} H_0 e^{ikx\sin\theta_{inc}} e^{-iky\cos\theta_{inc}} - i\frac{Z}{k} a_{sw} H_0 e^{-a_{sw}y}\cos(2kx\sin\theta_{trans}), \quad (59)$$

$$H_z^<(x,y) = H_0\sqrt{\frac{2\cos\theta_{inc}}{\cos\theta_{trans}}} e^{-iky\cos\theta_{trans}}\cos(kx\sin\theta_{trans}), \quad (60)$$

$$E_x^<(x,y) = Z\cos\theta_{trans} H_0\sqrt{\frac{2\cos\theta_{inc}}{\cos\theta_{trans}}} e^{-iky\cos\theta_{trans}}\cos(kx\sin\theta_{trans}), \quad (61)$$

where

$$a_{sw} = k\sqrt{4\sin^2\theta_{trans}-1}. \quad (62)$$

The $\cos(2kx\sin\theta_{trans})$ factor in Eqs. (58) and (59) represents the above-mentioned surface waves $\exp(i2kx\sin\theta_{trans})$ and $\exp(-i2kx\sin\theta_{trans})$ that are traveling in the $+x$ and $-x$ directions, respectively. These surface waves turn out to be the $\rho_{-2}$ and $\rho_2$ components of the Floquet-Bloch (FB) field expansion in $y>0$ (see Eq. (A4)) which represent evanescent waves when $a_{sw}>0$. The $\cos(kx\sin\theta_{trans})$ factor in Eqs. (60) and (61) represents the two desired split waves in the $+\theta_{trans}$ and $-\theta_{trans}$ directions. Using Eqs. (58) to (61) in Eqs. (10) and (11), and the result in Eqs. (3) to (5) will again yield the values of $K_{em}(x)$, $Z_{se}(x)$ and $Y_{sm}(x)$. Using these, in turn, in Eqs. (12) and (13) will provide $T(x)$ and $R(x)$. Since the splitting functionality is symmetric, it is expected that the $T(x)$ and $R(x)$ would be symmetric as well. This can be clearly seen relative to $x/d=0.5$ in Fig. 9 which displays $T(x)$ and $R(x)$ for the representative case $\theta_{trans}=80°$ that will be used in the remainder of this discussion.

The $T(x)$ and $R(x)$ as shown in Fig. 9 can be evaluated at each $x=x_p$ for use in Eqs. (18) or (19) to determine the widths $w_{pi}$ of the layers in each waveguide which defines the FP-OBMS structure. Fig. 10 displays a structure designed in this manner using the sample functions $T(x)$ and $R(x)$ displayed in Fig. 9.

The fields produced by a wave normally incident on the structure in Fig. 10 are shown in Fig. 11 as computed by CST (full-wave solutions), and analytically by solving Eqs. (24) and (25). The purely horizontal orientation of the wave-fronts in the upper region demonstrates the absence of a reflected wave, as designed. This was verified by CST which found that the power coupling efficiency to specular reflection was less than 1%. The field pattern in the lower region is, of course, due to the interference between the two split waves, the directions of which are indicated by sideward arrows in the lower region. CST indicated that the power coupling efficiency to the split waves was the same for each wave and very close to 50%. The agreement is excellent between the full-wave and the analytical method in Fig. 11, both outside and inside the FP-OBMS structure.

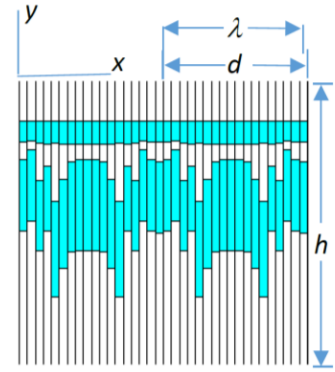

FIG. 10. Two periods of the FP-OBMS for beam splitting, $N_{wg}=18$, $\theta_{trans}=80°$, $d=1.0154\lambda$, structure obtained from Eqs. (18), $h=2\lambda$. This FP-OBMS produces the $T(x)$ and $R(x)$ of Fig. 9.

Also shown in Fig. 11 is the field for the "ideal" OBMS geometry involving a zero-thickness surface, and defined by Eqs. (58) and (60). For each case in Fig. 11, the field includes two FB reflection components $\rho_{-2}$ and $\rho_2$ which together produce a standing surface wave. This surface wave is indicated by vertical arrows in Fig. 11 which point to the field disturbances which it produces which are located just above the metasurface [18]. Remarkably, this surface wave which appears "by definition" in the field for the ideal zero-thickness metasurface is reproduced by the two-wavelength-thick FP-OBMS! This demonstrates the capability of the FP-OBMS to emulate the ideal zero-thickness OBMS in its entirety: both the propagating waves and the evanescent waves. To the best of our knowledge, this is the first time that a realistic OBMS structure has been designed with such a capability.

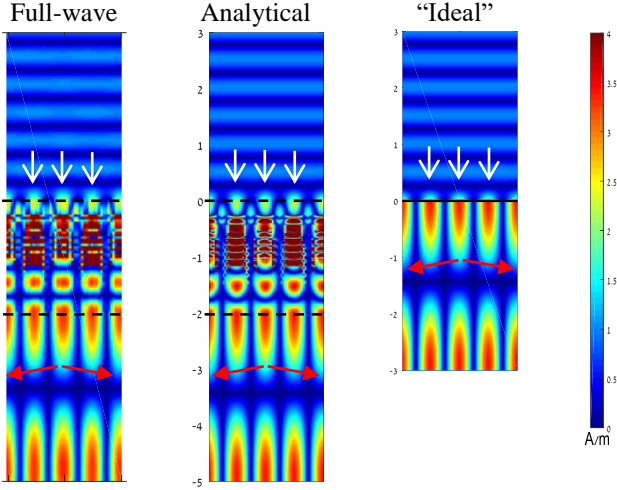

FIG. 11. H field color images of propagation through FP-OBMS structures of Fig. 10 which were designed to split a normally incident plane wave into two waves in symmetrically opposite directions. Full-wave solution, analytical results, and ideal (zero-thickness) results are shown for splitting angle $\theta_{trans}=80°$. Red oblique arrows indicate the directions of the split waves. Downward white vertical arrows point to the surface wave. All geometric scales are relative to a wavelength. The horizontal extent of each plot is two periods.

## VI. CONCLUSION

We have introduced the concept of a FP-OBMS for passively and losslessly molding an existing electromagnetic field into a desired new field. The only requirements on this new field are that it satisfy Maxwell's equations and that the power flow across the metasurface be locally continuous. The field that is produced by the thick FP-OBMS reliably reproduces the complete performance properties of the ideal (abstract) zero-thickness OBMS: not only does it emulate the propagating waves scattered by the corresponding zero-thickness OBMS, but it also emulates the auxiliary non-local excitation of evanescent waves required to achieve certain functionalities with high efficiency. This has been verified with the aid of analytical predictions and full-wave computations.

The thick *asymmetric* FP-OBMS design was implemented for anomalous refraction and for beam splitting using an array of parallel-plate waveguides containing two cascaded dielectric layers, the widths of which are adjusted in accordance with Fabry-Perot theory to provide proper reflection and transmission coefficients.

For OBMS anomalous refraction, a closed-form field expression was obtained which, when compared with the analogous HMS expression, provided insight into the underlying relationship between the two types of metasurfaces. Employing a geometrical optics model, it was seen that the OBMS effectively adds a virtual anti-reflective coating to the HMS which provides the sought-for perfect anomalous refraction. This interpretation is used to investigate and explain for the first time differences in the angular response of OBMS in comparison to HMS. It was shown that for minor wave-impedance mismatch, the HMS and OBMS angular responses are almost identical, whereas for wide-angle refraction, the virtual anti-reflective coating is required to produce a large "reflection coefficient" promoting a "resonant" behavior. This behavior results in an enhanced efficiency for the OBMS, but in a rather narrow angular range around the designated angle of incidence.

The capabilities demonstrated by the FP-OBMS pave the path to practical realization of wave transmission with intricate functionalities, including those which have never before been associated with a physical structure.

## APPENDIX

Assuming TM propagation and no field or geometry variation in the $z$ direction, the magnetic field can be written $\mathbf{H}(x,y) = H(x,y)\hat{\mathbf{z}}$, where $H(x,y)$ satisfies the Helmholtz equation

$$\frac{\partial^2 H(x,y)}{\partial x^2} + \frac{\partial^2 H(x,y)}{\partial y^2} + k^2 H(x,y) = 0. \quad (A1)$$

Referring to Fig. 1, the fields will have different expressions above, below and within the FP-OBMS:

$$H(x,y) = \begin{cases} H_{inc}(x,y) + H_{ref}(x,y), y > 0, \\ H_2(x,y), -h \leq y \leq 0, \\ H_{trans}(x,y), y < -h, \end{cases} \quad (A2)$$

where the incident wave of unit amplitude is given by

$$\mathbf{H}_{inc}(x,y) = \hat{\mathbf{z}} e^{ikx\sin\psi_{inc} - iky\cos\psi_{inc}}. \quad (A3)$$

The field $H_2(x,y)$ is defined in Eq. (21). The reflected field $H_{ref}$ and the transmitted field $H_{trans}$ are defined in accordance with Fourier-Bloch (FB) theory as superpositions of reflected plane waves and transmitted plane waves, respectively:

$$H_{ref} = \sum_{n=-\infty}^{\infty} \rho_n e^{i\alpha_n x + i\beta_n y}, \quad H_{trans} = \sum_{n=-\infty}^{\infty} \tau_n e^{i\alpha_n x - i\beta_n(y+h)}, \quad (A4)$$

where

$$\alpha_n = k\sin\psi_{inc} + \frac{2n\pi}{d}, \quad \beta_n = \sqrt{k^2 - \alpha_n^2}, \quad (A5)$$

$d$ is defined in Eq. (23), $\rho_n$ and $\tau_n$ are the initially unknown amplitudes of the reflected and transmitted waves, respectively, and the branch of $\beta_n$ is chosen to satisfy the radiation condition for $|y|\to\infty$.

With the aid of Eqs. (21) and (A2) to (A4), the conditions for the continuity of the $E$ and $H$ field components tangent to the upper ($y=0$) and lower ($y=-h$) boundaries of the FP-OBMS may be written:

$$H_{inc}(x,0) + H_{ref}(x,0) = \sigma_{pa} + T_p\sigma_{pb} + R_p\sigma_{pa}, \quad (A6)$$

$$\frac{\partial H_{inc}(x,0)}{\partial y} + \frac{\partial H_{ref}(x,0)}{\partial y} = -ik\sigma_{pa} + ik(T_p\sigma_{pb} + R_p\sigma_{pa}), \quad (A7)$$

$$H_{trans}(x,-h) = (T_p\sigma_{pa} + Q_p\sigma_{pb})e^{ikh} + \sigma_{pb}e^{-ikh}, \quad (A8)$$

$$\frac{\partial H_{trans}(x,-h)}{\partial y} = -ik(T_p\sigma_{pa} + Q_p\sigma_{pb})e^{ikh} + ik\sigma_{pb}e^{-ikh}. \quad (A9)$$

These four equations can be appreciably simplified by solving for $\sigma_{pa}$ and $\sigma_{pb}$. From Eqs. (A6) and (A7), it is easily found that

$$\sigma_{pa} = \frac{1}{2}\left[H_{inc}(x,0) + H_{ref}(x,0) - \frac{1}{ik}\frac{\partial H_{inc}(x,0)}{\partial y} - \frac{1}{ik}\frac{\partial H_{ref}(x,0)}{\partial y}\right]. \quad (A10)$$

Similarly, from Eqs. (A8) and (A9),

$$\sigma_{pb} = \frac{e^{ikh}}{2}\left[H_{trans}(x,-h) + \frac{1}{ik}\frac{\partial H_{trans}(x,-h)}{\partial y}\right]. \quad (A11)$$

Substituting Eqs. (A10) and (A11) in Eqs. (A6) and (A8) produces:

$$-\frac{1}{2}\left[(1-R_p)H_{ref}(x,0) + (1+R_p)\frac{1}{ik}\frac{\partial H_{ref}(x,0)}{\partial y}\right]$$
$$+T_p\frac{e^{ikh}}{2}\left[H_{trans}(x,-h) + \frac{1}{ik}\frac{\partial H_{trans}(x,-h)}{\partial y}\right]$$
$$= \frac{1}{2}\left[(1-R_p)H_{inc}(x,0) + (1+R_p)\frac{1}{ik}\frac{\partial H_{inc}(x,0)}{\partial y}\right], \quad (A12)$$

$$\frac{1}{2}\left[-(1-Q_pe^{2ikh})H_{trans}(x,-h) + (1+Q_pe^{2ikh})\frac{1}{ik}\frac{\partial H_{trans}(x,-h)}{\partial y}\right]$$
$$+T_p\frac{1}{2}\left[H_{ref}(x,0) - \frac{1}{ik}\frac{\partial H_{ref}(x,0)}{\partial y}\right]e^{ikh}$$
$$= -T_p\frac{1}{2}\left[H_{inc}(x,0) - \frac{1}{ik}\frac{\partial H_{inc}(x,0)}{\partial y}\right]e^{ikh}. \quad (A13)$$

Using Eqs. (A3) to (A5) in Eqs. (A10) to (A13) yields

$$\sigma_{pa} = \sigma_a(x) = C_0e^{i\alpha_0 x} + \sum_{n=-\infty}^{\infty} S_n\rho_n e^{i\alpha_n x}, \quad (A14)$$

$$\sigma_{pb} = \sigma_b(x) = e^{ikh}\sum_{n=-\infty}^{\infty} S_n\tau_n e^{i\alpha_n x}, \quad (A15)$$

$$-\sum_{n=-\infty}^{\infty} \rho_n(C_n - R_pS_n)e^{i\left(\frac{2n\pi}{d}\right)x} + T_p e^{ikh}\sum_{n=-\infty}^{\infty} \tau_n S_n e^{i\left(\frac{2n\pi}{d}\right)x}$$
$$= S_0 - R_pC_0 \quad (A16)$$

$$-T_p\sum_{n=-\infty}^{\infty} \rho_n S_n e^{i\left(\frac{2n\pi}{d}\right)x} + e^{-ikh}\sum_{n=-\infty}^{\infty} \tau_n(C_n - Q_pe^{2ikh}S_n)e^{i\left(\frac{2n\pi}{d}\right)x}$$
$$= T_pC_0, \quad (A17)$$

where $1 \leq p \leq N_{wg}$,

$$S_n = \frac{1}{2}(1-\gamma_n), \quad C_n = \frac{1}{2}(1+\gamma_n), \quad \gamma_n = \frac{\beta_n}{k} \equiv \cos\psi_n. \quad (A18)$$

From Eqs. (A18) and (A5), the following identities are useful:

$$C_0 = \cos^2\left(\frac{\psi_{inc}}{2}\right), \quad S_0 = \sin^2\left(\frac{\psi_{inc}}{2}\right), \quad (A19)$$

$$C_{-1} = \cos^2\left(\frac{\psi_{trans}}{2}\right), \quad S_{-1} = \sin^2\left(\frac{\psi_{trans}}{2}\right). \quad (A20)$$

The discrete functions $T_p$, $R_p$, $Q_p$ will now be approximated as continuous functions $T(x)$, $R(x)$, $Q(x)$ which, as shown in Eq. (22) can be written as Fourier series. When these are substituted into Eqs. (A16) and (A17), multiple summations appear, and may be rearranged using

$$\sum_{j=-\infty}^{\infty} a_j e^{i\left(\frac{2j\pi}{d}\right)x} \sum_{n=-\infty}^{\infty} b_n e^{i\left(\frac{2n\pi}{d}\right)x} = \sum_{n=-\infty}^{\infty} c_n e^{i\left(\frac{2n\pi}{d}\right)x}, \quad (A21)$$

where

$$c_n = \sum_{j=-\infty}^{\infty} a_{n-j}b_j \quad (A22)$$

Then Eqs. (A16) and (A17) may be written

$$-\sum_{n=-\infty}^{\infty}\left(\rho_n C_n - \sum_{j=-\infty}^{\infty} a_{r(n-j)}\rho_j S_j\right)e^{i\left(\frac{2n\pi}{d}\right)x}$$
$$+e^{ikh}\sum_{n=-\infty}^{\infty}\sum_{j=-\infty}^{\infty} a_{t(n-j)}\tau_j S_j e^{i\left(\frac{2n\pi}{d}\right)x}$$
$$= \sum_{n=-\infty}^{\infty} [\delta_{n0}S_0 - C_0 a_{rn}]e^{i\left(\frac{2n\pi}{d}\right)x}, \quad (A23)$$

$$-\sum_{n=-\infty}^{\infty}\sum_{j=-\infty}^{\infty} a_{t(n-j)}\rho_j S_j e^{i\left(\frac{2n\pi}{d}\right)x}$$

$$+e^{-ikh}\sum_{n=-\infty}^{\infty}\left(\tau_n C_n - e^{2ikh}\sum_{j=-\infty}^{\infty} a_{q(n-j)}\tau_j S_j\right)e^{i\left(\frac{2n\pi}{d}\right)x}$$

$$= C_0 \sum_{n=-\infty}^{\infty} a_{tm} e^{i\left(\frac{2n\pi}{d}\right)x}, \qquad (A24)$$

Equating exponential coefficients of *x*:

$$-\left(\rho_m C_m - \sum_{j=-\infty}^{\infty} a_{r(m-j)}\rho_j S_j\right) + e^{ikh}\sum_{j=-\infty}^{\infty} a_{t(m-j)}\tau_j S_j$$

$$= \delta_{m0} S_0 - a_{rm} C_0, \qquad (A25)$$

$$-\sum_{j=-\infty}^{\infty} a_{t(m-j)}\rho_j S_j + e^{-ikh}\left(\tau_m C_m - e^{2ikh}\sum_{j=-\infty}^{\infty} a_{q(m-j)}\tau_j S_j\right)$$

$$= a_{tm} C_0, \qquad (A26)$$

for all positive and negative values of the index *m*. The truncated form of these equations are given in Eqs. (24) and (25).

---